\documentclass[pra, twoside, twocolumn, showpacs]{revtex4}

\usepackage{amsmath,amsfonts,amssymb,epsfig}

\newcommand{\bra}[1]{\ensuremath{\langle#1|}}
\newcommand{\ket}[1]{\ensuremath{|#1\rangle}}
\newcommand{\braket}[2]{\ensuremath{\langle#1|#2\rangle}}
\newcommand{\hilb}{\ensuremath{\mathcal{H}}}
\newcommand{\bbz}{\ensuremath{\mathbb{Z}}}
\newcommand{\bbc}{\ensuremath{\mathbb{C}}}
\newcommand{\ie}{{\frenchspacing i.e.~}}
\newcommand{\eg}{{\frenchspacing e.g.~}}
\newcommand{\pcnot}{\ensuremath{\phi\text{CNOT}}~}

\newcommand{\addfigure}[3][.8]{
  \begin{figure}[htb]
    \begin{center}
      \epsfig{height=#1\columnwidth,file=#2.eps}
    \end{center}
    \caption{#3}
    \label{fig:#2}
  \end{figure}
}

\begin{document}

\title{Scattering model for quantum random walks on hypercube}
\author{Jozef Ko\v s\'\i k$^{1}$ and Vladim\' \i r Bu\v{z}ek$^{1,2}$}
\address{$^{1}$Research Center for Quantum Information, Institute of Physics, Slovak Academy of Sciences,
  D\'{u}bravsk\'{a} cesta 9, 845 11 Bratislava, Slovakia\\
  $^{2}$Faculty of Informatics, Masaryk University, Botanick\'a 68a,
  602 00 Brno, Czech Republic}
\date{16 March 2004}
\pacs{03.67.-a, 05.40.Fb}

\begin{abstract}
  Following a recent work by M. Hillery {\it et al.} [ Phys. Rev. A {\bf 68}, 032314 (2003)] we
  introduce a scattering model of a quantum random walk (SQRW) on a
  hybercube. We show that this type of quantum random walk can be
  reduced to the quantum random walk on the line and we derive the
  corresponding hitting amplitudes. We investigate the scattering
  properties of the hypercube, connected to the semi-infinite tails. We prove
  that the SQRW is a generalized version of the coined quantum
  random walk. We show how to implement the SQRW efficiently using a
  quantum circuit with standard gates. We discuss one possible
  version of a quantum search algorithm using the SQRW. Finally we analyze
  symmetries that underlie the SQRW and may simplify its solution
  considerably.
\end{abstract}

\maketitle

\section{Introduction}

Quantum random walk is a theoretical concept conceived to simulate
certain algorithms using quantum mechanical elements, \ie unitary
operators and measurements \cite{kempe2003}. In particular, it has been shown recently that
it is possible to use a quantum random walk to perform a search in a
database with the topology of the  hypercube faster that it
can be done classically \cite{shenvi2003}. It is an oracle based
algorithm, which is optimal in its speed.
Another successful application of quantum
random walks has been demonstrated by Childs, {\em et al.} cite{childs2003} who
have also constructed an oracle problem that can be solved by a quantum algorithm
exploiting quantum random walk exponentially faster than any classical algorithm.

These two examples justify the general hope that quantum random walks
might be able to solve some problems, based
on random processes \cite{footnote1} (\eg Monte Carlo methods, 2-SAT,
graph connectivity, etc.) faster than corresponding classical
algorithms.  

In general, there are essentially three types  of quantum random
walks. Firstly, let us mention the so-called coined quantum random
walk (CQRW), that is a discrete time walk which makes use of 
an additional quantum system, the coin \cite{aharonov1993}. The second
type of the quantum random walks 
is described by a continuous (Hamiltonian) dynamics of a quantum
system \cite{childs2002}. The third type 
of quantum random walks based on physical model of optical multiports
has been recently been introduced by M. Hillery {\it et al.} \cite{hillery2003,feldman2004}.

Before we proceed we note, that
a quantum random walk as discussed by D. Aharonov {\it et al.}
\cite{aharonov2001} is basically a mapping 
$\psi:\mathcal{G_V}\rightarrow\bbc^d$, which is updated at each step by
a function $\psi(x)\mapsto F[\psi(y):(xy)\in\mathcal{G_E}]$, where
$\mathcal{G}=(\mathcal{G_V},\mathcal{G_E})$ is a graph with vertices
$\mathcal{G_V}$ and edges $\mathcal{G_E}$.

We therefore can say that quantum random walk is
a special instance of the quantum cellular automaton \cite{meyer1996, werner2003}.
The classical cellular automaton  is a concept general enough to
accommodate virtually any algorithm; more precisely, any Turing machine
can be simulated using a cellular automaton.

The CQRW is usually defined on regular graphs (each vertex having the
same number of outgoing edges). The definition on non-regular graphs
is also possible, and some interesting algorithms are based on this
version \cite{eisenberg2003}. However, the latter version does not
possess the symmetries of the former one, nor its neat tensor product
structure (the unitary evolution operator CQRW on the regular Cayley graph
commute with generators of the underlying group). Instead, the whole
graph must be addressed, by means of an oracle which tells us whether
any two vertices are connected by an edge \cite{kendon2003}, which
causes a considerable growth of the resources.

In this paper we will focus our attention on a quantum-optical model of multiports
\cite{hillery2003,feldman2004} which describes  a possible physical
realization of specific quantum random walks.
In this scheme, we have an
array of multiports (see e.g. \cite{jex1995} and references therein),
interconnected with optical paths. 
A photon is launched  into one path and is transformed by the action
of the multiports. This action can be described as a scattering process, therefore,
we will refer to this scheme as the scattering quantum random walk (SQRW).

The SQRW is more viable from the experimental point of view, can be
extended to non-regular graphs and is equivalent to CQRW on the
regular graphs.

We will investigate a particular arrangement of the multiports when
are localized at the vertices of a hypercube.
The array of multiports effectively acts as a
scattering potential, when connected to semi-infinite tails. It can be
endowed with two characteristic values, the reflection and
transmission amplitude for photons. This was done in Ref.~\cite{hillery2003,feldman2004}
for a special two-dimensional hypercube.

Our paper is organized as follows: in Sec. \ref{sec:definition} we
define the SQRW on the hypercube. In Sec. \ref{sec:topology} we
show how the SQRW on the hypercube may be reduced to the cellular
automaton on the line. In addition we will compute the hitting amplitude and we will make some
other simulations.
In Sec. \ref{sec:potential} we will investigate
scattering properties of the hypercube connected to
semi-infinite tails.  Sec. \ref{sec:superset} contains the proof
that the SQRW is equivalent to a generalized version of the  coined
quantum random walk. In Sec. \ref{sec:implementation} we show how
to implement the SQRW efficiently on the quantum circuit.
In Sec. \ref{sec:search} we discuss one possible
version of search algorithm using the SQRW. Finally, in Sec.
\ref{sec:symmetries}, we will analyze  symmetries that underlie the
SQRW and which may considerably simplify the solution of the model.

\section{The definition of SQRW}
\label{sec:definition}

The SQRW was first presented in \cite{hillery2003}.
The technique behind its implementation are the  multiports \cite{jex1995}: linear
optical elements, interconnected with optical paths. Each multiport
has in general different number of {\em inputs} (which at the same time serve also as outputs).
A coherent superposition of photons entering  the
multiport  is transformed into another coherent superposition of photons
outgoing from the multiport. The multiports are the vertices of a
graph $\mathcal{G_V}$, and the optical paths are its edges.

The photon travelling between multiports $x,y$
is denoted $\ket{xy}$. The Hilbert space on which the state of the
photon is defined is spanned by vectors
$\ket{xy},\;(xy)\in\mathcal{G_E}$ and can be decomposed into direct sum
of Hilbert subspaces $\hilb=\bigoplus_{x\in\mathcal{G_V}}\hilb_x$ where
$\hilb_x=\text{span}\{\ket{yx}:(yx)\in\mathcal{G_E}\}$. That is,
the base of $\hilb_x$ are the states of photon ingoing into multiport
at $x$. For convenience we introduce the Hilbert subspace spanned by
the states of photon outgoing from the vertex at $x$:
$\widehat\hilb_x=\text{span}\{\ket{xy}:(xy)\in\mathcal{G_E}\}$

The evolution operator of states in $\hilb$ is
$U=\bigoplus_{x\in\mathcal{G_V}}U_x$ where $U_x$ is isometry
$\hilb_x\rightarrow\widehat\hilb_x$ (onto, hence linear, hence
unitary). Since the subspaces are orthogonal, $U$ is unitary.

The concrete realization of unitary operator $U_x$ reflects the fact
that the multiport partially reflects and partially transmits the
ingoing photon. Denoting the reflection and transmission coefficients
respectively $r$ and $t$ we have
\begin{equation}
  U_x\ket{yx}=r\ket{xy}+t\sum_{(xz)\in\mathcal{G_E}}\ket{xz}
\end{equation}
where $(xy)\in\mathcal{G_E}$
For $U$ to be unitary, these coefficients must satisfy the relations \cite{hillery2003}
\begin{eqnarray}
  \label{eq:cond_rt}
  |r|^2+(d-1)|t|^2&=&1\\
  (d-2)|t|^2+r^*t+rt^*&=&0
\end{eqnarray}
The operator $U_x$ (for any $x$) has in the natural basis the matrix
the form $(U_x)_{ij}=r\delta_{ij}+t(1-\delta_{ij})$. The relations
(\ref{eq:cond_rt}) may be satisfied by setting by using the Grover coefficients
$r=\frac{2}{d}-1,\,t=\frac{2}{d}$, see \cite{moore2001}.
The pseudo-eigensystem of any such $U_x$ can be readily computed
(``pseudo'' means that we set an isomorphism between $\hilb_x$ and
$\widehat\hilb_x$ s.t. $\ket{yx}\equiv\ket{xy}$).
Since
$U_x=2\ket{s}\bra{s}-1$ ($\ket{s}$ is the complete superposition over
the basis of the domain of $U_x$, which we denote
$\ket{1},\dots,\ket{d}$),
the eigenvectors are $\ket{s}$ (with pseudo-eigenvalue 1)
and linearly independent (but not orthogonal) set
$\{\ket{1}-\ket{k}:k=2,\dots,d\}$ (with 
$(d-1)$-degenerate pseudo-eigenvalue -1). Making the direct sum of these
eigenvectors gives us a pseudo-eigensystem of $U$, with
pseudo-eigenvalues $\pm1$.

There are many other choices of the reflection and the transmission coefficients,
one set of them is the following (we will use it when necessary):
\begin{equation}
  \label{eq:rt}
  \begin{split}
    t&=\frac{1}{d^p}\\
    r&=\sqrt{1-\frac{d-1}{d^{2p}}}\;e^{i\theta}
  \end{split}
\end{equation}
where $\cos\theta=\frac{1-d/2}{\sqrt{d^{2p}-d+1}}$ with $p>1/2 $.
If we set $p=1$, then the reflection  amplitude goes to
$\lim_{d\rightarrow\infty} r=-\frac{1}{2}+i\frac{\sqrt3}{2}$.
Obviously, for large dimensions, almost all of  photons will be
reflected. The multiports with these coefficients will be called
{\em symmetric} multiports.
We will later compare both sets of coefficients with
respect to their mixing properties.

From now on we will be dealing with the multiports arranged into the
form of the $d$ dimensional hypercube. The edges of the hypercube will
form  two-way optical paths.

The $d$-dimensional hypercube is the Cayley graph
$\mathcal{G}=(\bbz_2^d,[d])$ where $[d]$ is the set of generators of
the additive (mod 2) group $\bbz_2^d$ (the binary strings with only
one 1).
For any $x,y\in\bbz_2^d$ we set the scalar product
$xy=x_1y_1+\dots+x_dy_d$ (mod 2). The norm $|x|=\sqrt{xx}$ is the
Hamming weight (the number of 1s in $x$). The set $\ell_w=\{x:|x|=w\}$
is called the \textsl{layer} of the hypercube. Obviously $[d]=\ell_1$.

\addfigure{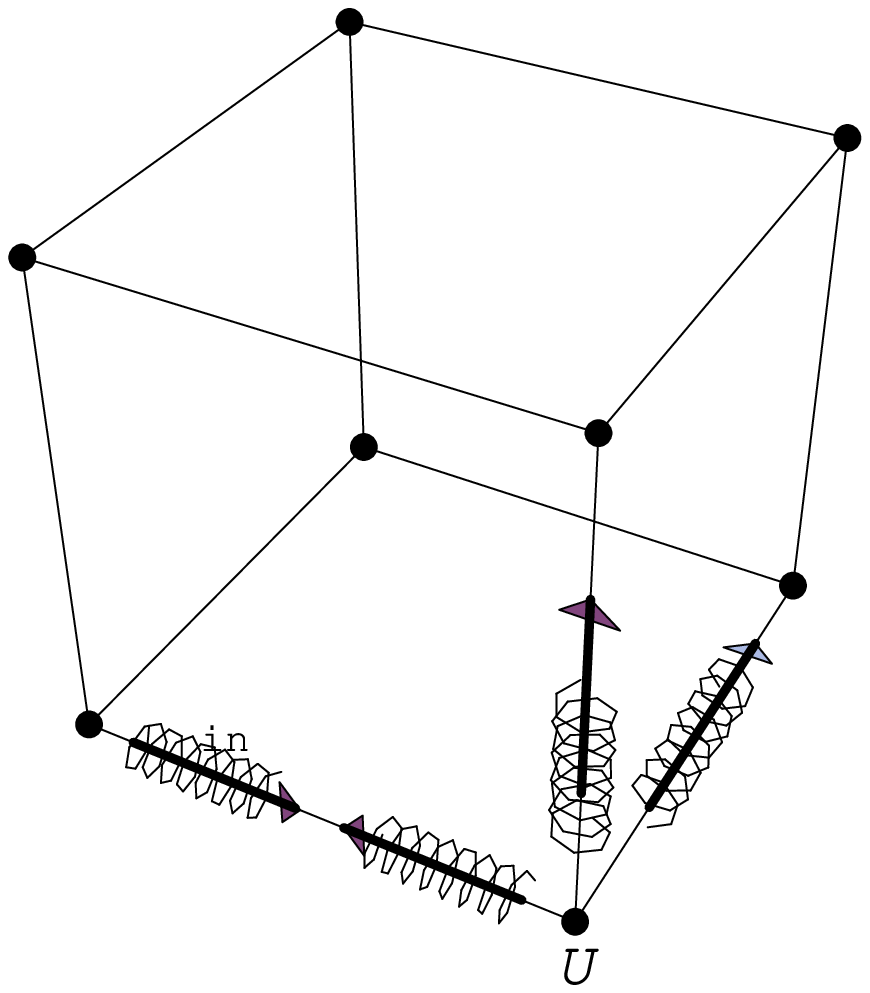}{Action of the multiport ($U$) on the ingoing
  photon (in) on the dim 3 hypercube. Coherent superposition of 3
  photonic excitations is created.}

For simplicity, we will denote the basis states of $\hilb$ for the
$d$-hypercube $\ket{xy}$ like $\ket{x;a}$ where $x$ is the vertex and
$a=1,\dots,d$ is the generator s.t. $x+a=y$.

\section{The topology of the hypercube}
\label{sec:topology}

The hypercube may be broken up in individual layers $\ell_w$, \ie sets of
vertices with equal Hamming weight. There is a special class of
vectors from $\hilb$, which is closed w.r.t $U$ and whose members may
be described by fewer coefficients, thus simplifying the evolution
equations. Namely, these are the vectors $\ket{\psi}$
s.t. $\braket{x;a}{\psi}$ is the same for all $|x|=w$ and for all
$a$. Under this assumption, the vectors are specified by coefficients
$\{\psi_{w,\pm}\}$ where $\psi_{w,\pm}=\braket{x;a}{\psi}$ with
$|x|=w,|x+a|=w\pm1$.

The reduced equations for evolution of the coefficients
$\psi_{w,\pm}$ are given from the assumptions that each vertex from
$\ell_w$ has a fixed number of edges which connect it to the previous
and next layer (with Hamming weight having $w\pm1$). We note that
a vertex $|x|=w$ is connected with  $w$ edges from previous layer
and with $d-w$ edges with next layer. Since the
coefficients assigned to the projection $\ket{\psi}$ to a given layer
are fixed, we get the recursive relations:
\begin{eqnarray}
  \label{eq:alphas}
  \begin{split}
    (U\psi)_{w,+}&=tw\psi_{w-1,+}+[t(d-w-1)+r]\psi_{w+1,-}\\
    (U\psi)_{w,-}&=t(d-w)\psi_{w+1,-}+[t(w-1)+r]\psi_{w-1,+}\\
  \end{split}
\end{eqnarray}
For $w=0,d$ these equations still hold wherever it makes sense, \ie
for $\psi_{0,+},\psi_{d,-}$. The coefficients
$\psi_{-1,\cdot},\psi_{d+1,\cdot}$ are neglected as long as they are
multiplied by zero in the equation.

The evolution which is governed by these equations is called the
symmetric SQRW.

If the initial state is $\psi_{0,+}=\frac{1}{\sqrt{d}}$,
then we immediately obtain the expression for the hitting amplitude 
\begin{equation}
  \label{eq:hit_ampl}
  \psi_{d,-}(d)=[t(d-1)+r]\;(d-1)!\;\frac{t^{d-1}}{\sqrt{d}}
\end{equation}
Using the Stirling approximation $\log n!\approx n\log n -n$ we can express
for large $d$ the above equation as $\psi_{d,-}(d)= \theta(e^{-d})$. We
see that the hitting amplitude drops exponentially with
$d$. Classically, the probability that we are at a given vertex from
$\ell_w$, is $p_w$, for which holds
$(Wp)_w=\frac{1}{d}wp_{w-1}+\frac{1}{d}(d-w)p_{w+1}-1$. From this we
get the hitting probability $(W^dp_0)=\frac{1}{d^d}d!$. By simple means
we show that  quantum and classical probabilities to get
from the vertex $|x|=0$ to vertex $|x|=d$ in $d$ steps are 
\begin{equation}
  \begin{split}
    p_c&=\frac{d!}{d^d}=\theta(e^{-d})\\
    p_q&=|\psi_{d,-}(d)|^2=\theta\left(\frac{(e/2)^{-d}}{\sqrt{d}}\right)
  \end{split}
\end{equation}
The classical and quantum hitting probabilities are related like
\begin{equation*}
  p_q=\theta\left(\frac{p_c^{1-\log2}}{\sqrt{d}}\right)
\end{equation*}
The ratio $\frac{p_q}{p_c}$ related to the dimension of the hypercube
is given on the figure Fig.\ref{fig:ratio}.
\addfigure[.6]{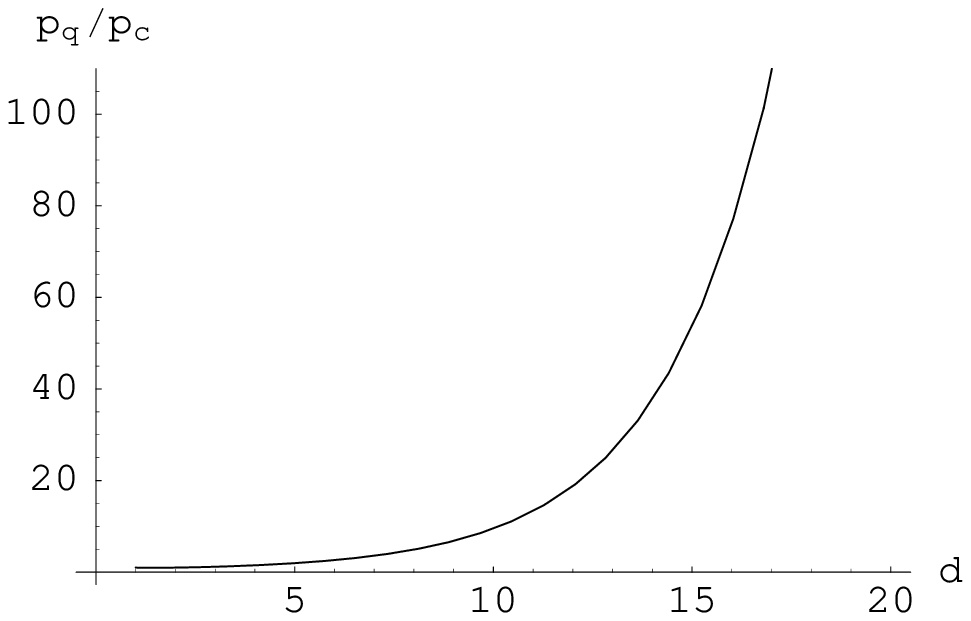}{The ratio of hitting probability for classical and
  quantum random walk on the hypercube, related to the dimension $d$.}
The whole model can be used to simulate coined quantum random walk on
the segment of line with coin dependent on the position. The state of
the coin is described by vector $[\psi_{w,+},\psi_{w,-}],w=0,\dots,d$ and is
updated at each step with update rules (\ref{eq:alphas}).
We might be troubled by the
fact that the update rule is not unitary. But we need the unitarity
only to conserve the inner product $\braket{\vec\alpha}{\vec\alpha}$.
Actually, the inner product
\begin{equation}
  \label{eq:inner_product}
  \braket{\vec\psi}{\vec\psi}=\sum_{w=0}^d{\binom{d}{w}}^2(|\psi_{w,+}|^2+
  |\psi_{w,-}|^2) 
\end{equation}
is conserved.

Though we have simplified the problem by the assumption  of symmetric
initial values, we are still far from its explicit solution. The
solution would rely on the path integration along different
paths by which two sites can be connected in a presupposed number
of steps. Each path would be assigned a complex amplitude (basically
some product of $r,t$), and by adding all the relevant paths together,
we would get the amplitude distribution over the hypercube. This
normally gives us enormous combinatorial expressions, which are
difficult to interpret.

Since the SQRW on the hypercube with symmetric initial states is
equivalent with one-dimensional (quantum) random walk on the finite
sequence of layers $\ell_w$, we may explore the probability
distribution $p_n(w)$ over the layers for an initial state
$\ket{\psi_0}=\sum_a\frac{1}{\sqrt{d}}\ket{0\dots0;a}$, where $p_n(w)$ is
the expectation value of the operator
\begin{equation}
  \label{eq:projop}
  M_w:=\sum_{x\in\ell_w,a}\ket{x;a}\bra{x;a}
\end{equation}
\ie $p_n(w)=\bra{\psi_0}(U^\dagger)^nM_wU^n\ket{\psi_0}$. We simulate
the evolution of $p_n(w)$; see Fig.\ref{fig:probdist}.

\addfigure[.55]{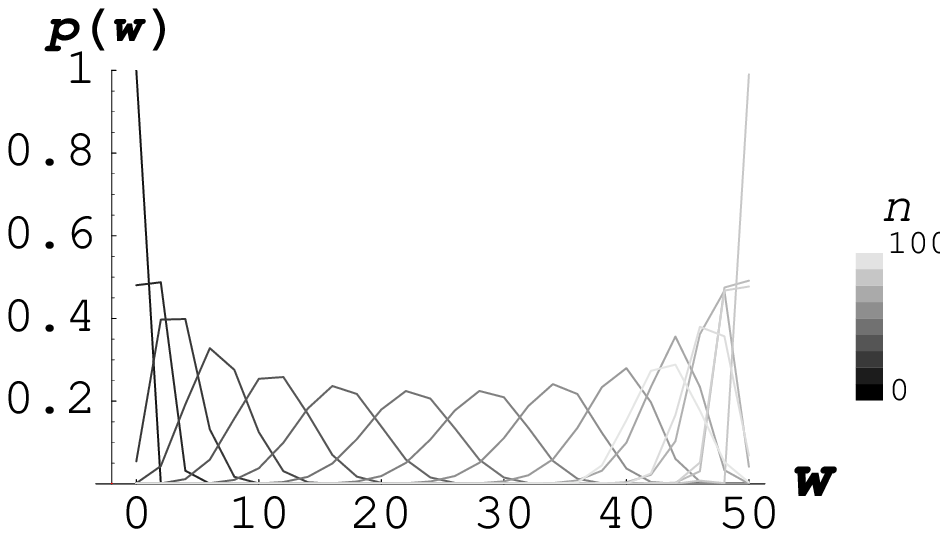}{The probability distribution of the SQRW,
  $p_n(w)$, for the hypercube of the dimension $d=50$, with a symmetric initial
  state localized at the vertex $0\dots0$, and steps
  $n=1,\dots,100$. The multiports are Grover.}

We see that the SQRW on the hypercube is effectively isomorphic to a dynamics of a
resonator: we start with one excited site, and the excitation
propagates as a Gaussian packet along the resonator. As the packet hits the boundary, the corresponding
site is excited, and the packet is reflected in the opposite
direction.

\addfigure[.7]{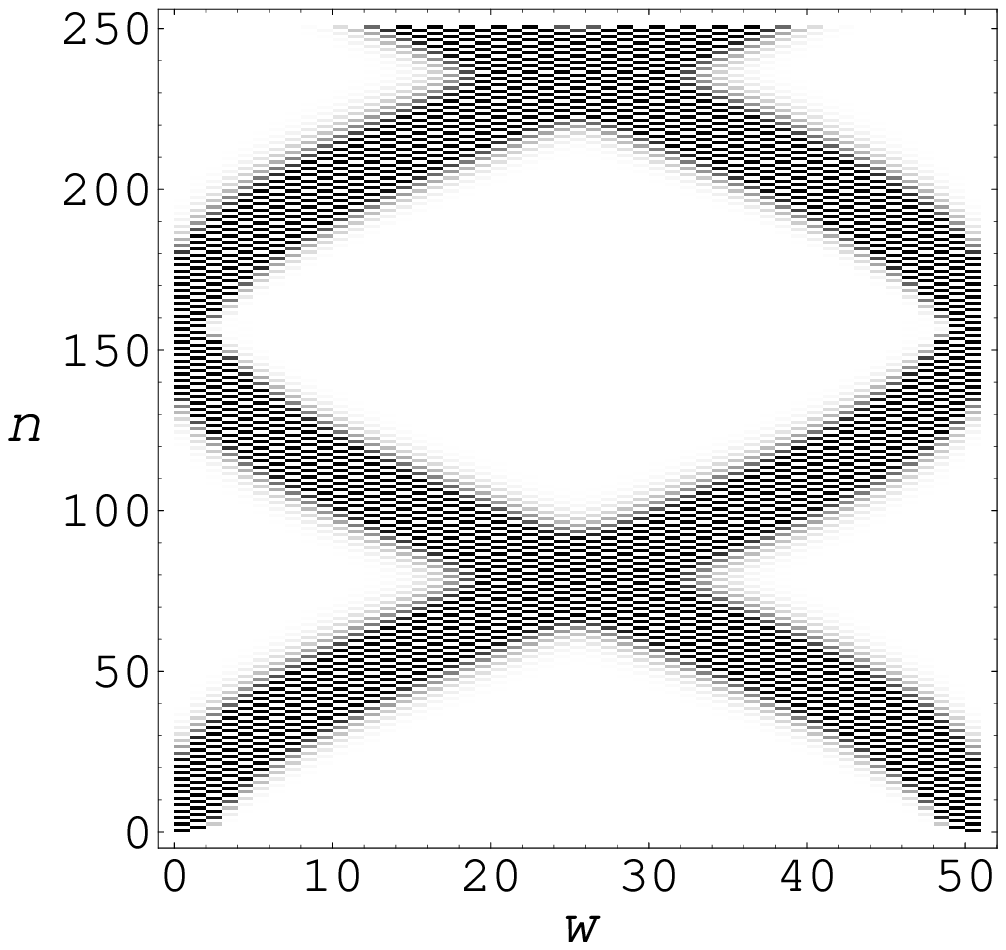}{The probability distribution of the
  symmetric SQRW on the hypercube of dim 50, for steps $n$ 0 to 250,
  with the initial state $\psi_{0,+}=\psi_{d,-}=\frac{1}{\sqrt{2d}}$
  (other $\psi$-s=0) and symmetric  multiports.}
\addfigure[.7]{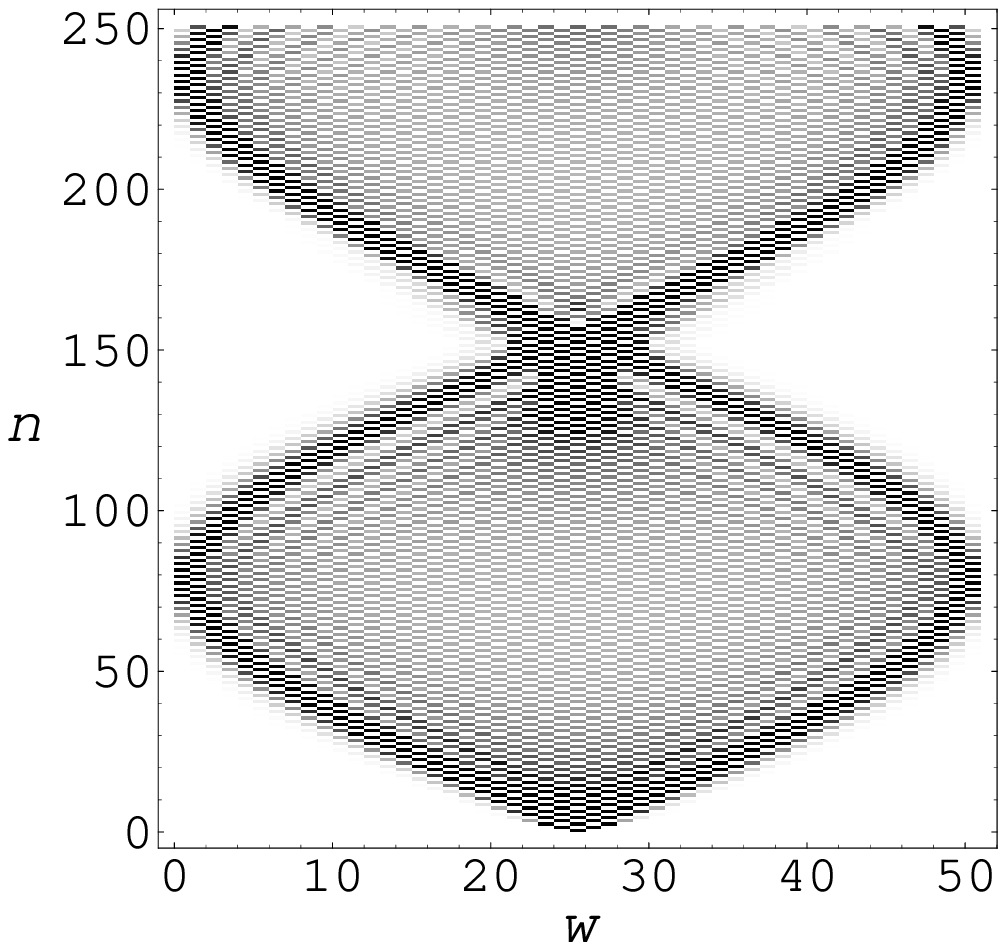}{The probility distribution of the
  symmetric SQRW on the hypercube of dim 50, for steps $n$ 0 to 250,
  with the initial state $\psi_{d/2+1,\pm}=1/\sqrt{2\binom{d}{d/2+1}}$
  (other $\psi$-s=0) and symmetric  multiports.}

\addfigure[.7]{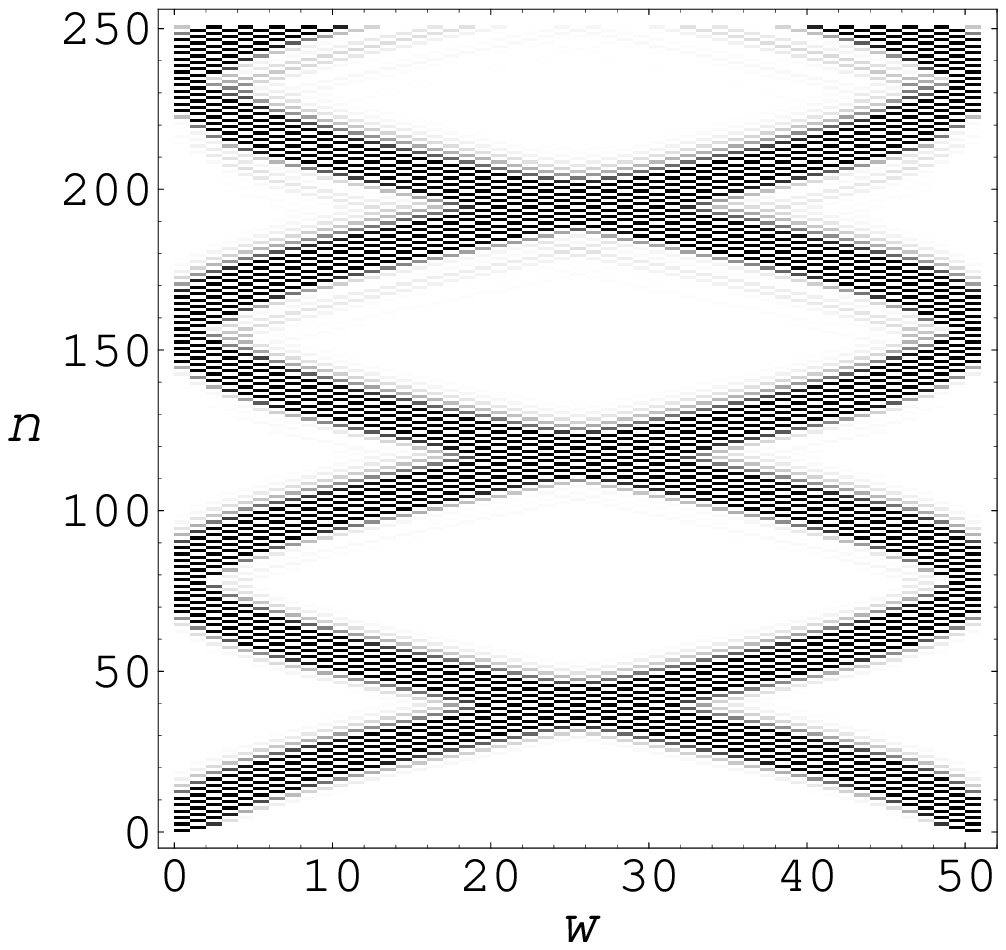}{The probability distribution of the
  symmetric SQRW on the hypercube of dim 50, for steps $n$ 0 to 250,
  with the initial state $\psi_{0,+}=\psi_{d,-}=\frac{1}{\sqrt{2d}}$
  (other $\psi$-s=0) and Grover multiports).}
\addfigure[.8]{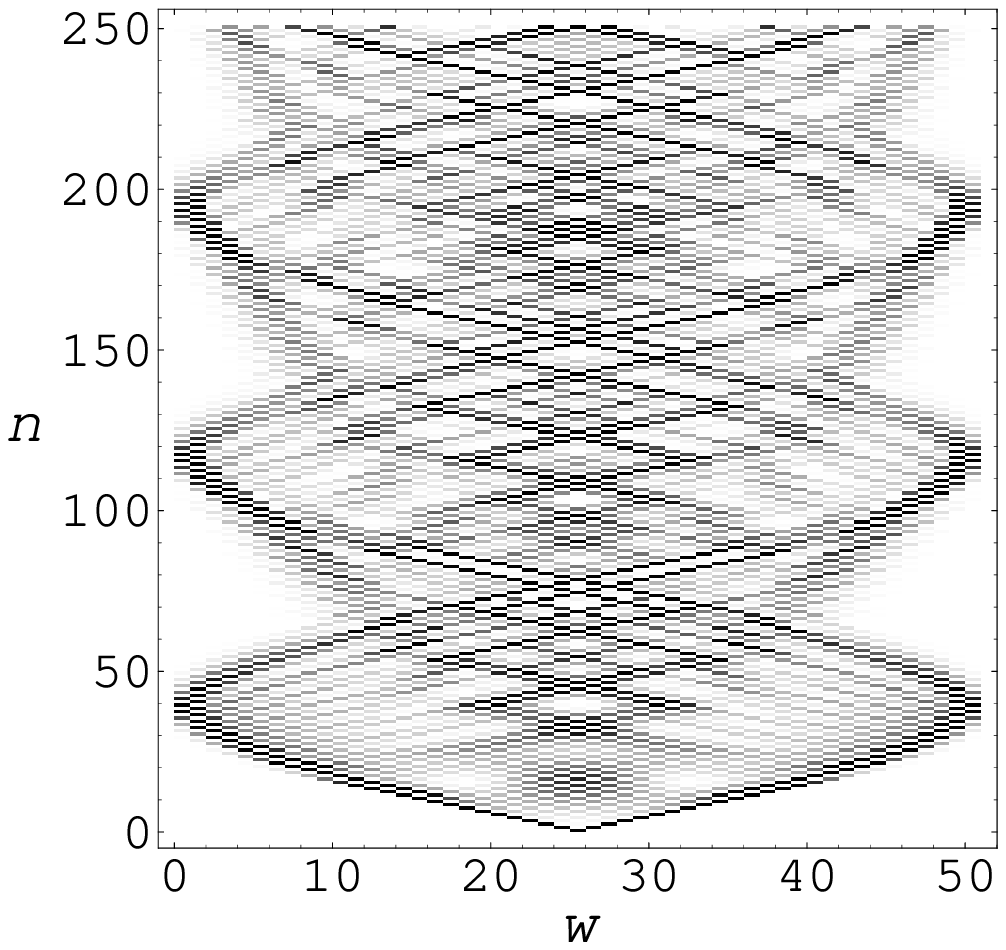}{The probability distribution of the
  symmetric SQRW on the hypercube of dim 50, for steps $n$ 0 to 250,
  with the initial state $\psi_{d/2+1,\pm}=1/\sqrt{2\binom{d}{d/2+1}}$
  (other $\psi$-s=0) and Grover  multiports.}

From Fig. \ref{fig:probdist} we see  that the hypercube acts as a
resonator when it comes to the evolution of  the probability
distribution over the layers.

We have performed
detailed simulations  for the symmetric SQRW with initial
conditions such that either only the coefficients
$\psi_{0,+};\, \psi_{d,-}$ are non-zero, or
$\psi_{d/2,\pm}\neq0$. There are two choices of the multiports: the
Grover and symmetric multiports. We make the simulations for both of them,
and for both initial conditions
(see Figs. \ref{fig:probdist2}--\ref{fig:probdist3g}). We see the
periodicity of the evolution: another feature akin to the resonant
behavior. Note that the
Grover multiport has much better mixing properties than the symmetric
multiport, owing to the fact that it is more distant from the unity.

\section{Hypercube as a scattering potential}
\label{sec:potential}

In Ref.~\cite{hillery2003} a model of 2-d hypercube with
semi-infinite tails attached to the vertices
with Hamming weights 0 and 2 has been studied.
Each tail has been supposed to be  a
1-d lattice  with perfectly transmitting multiports. Along one tail, a
photon enters the hypercube, and emerges on the other side. It is
possible to calculate explicitly the transmission coefficient of the
whole structure.
In the present section we will study analogous problem for an arbitrary-dimensional hypercube.
We will utilize some symmetry assumptions,
which let us perform the calculation (or at least the simulation) for arbitrary
higher dimensions.

The scheme we consider is shown on the figure \ref{fig:potential}.

\addfigure[1]{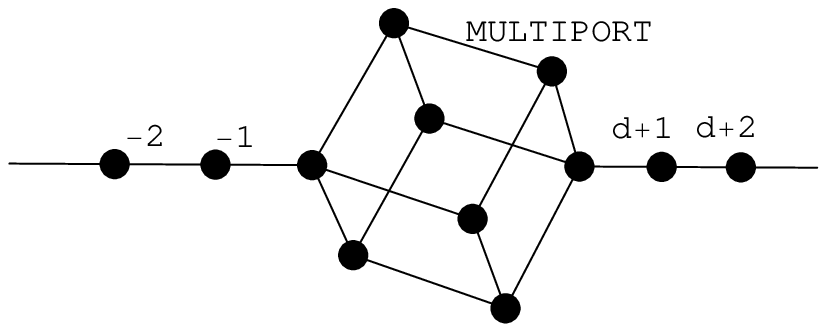}{The scattering potential (hypercube dim 3). The
  vertices outside the hypercube are denoted $-\infty,\dots,-2$ and
  $d+1,\dots$ for hypercube of dimension $d$.}

Now everything is as before, except that the multiports at the
vertices with Hamming weight $0$ and $d$ have reflection/transmission
coefficients $\tilde{r}=\frac{2}{d+1}-1\;\tilde{t}=\frac{2}{d+1}$. The
multiports outside the hypercube are perfectly transmitting.
The initial state is a
photon travelling between vertex $-1$ to vertex $|x|=0$ from the
hypercube.
The state of the whole system is described
by complex numbers $\alpha_{w,\pm}$ which represent the amplitude that
the photon travels from a vertex $|x|=w$ and is directed to the next or
previous layer $\ell_{w\pm1}$. Now we let $w$ running through $\bbz$.
The resulting relations are:
\begin{equation}
  \label{eq:alphas_sd}
  \begin{split}
    (U\psi)_{0,+}&=\tilde{t}\psi_{-1,+}+[(d-1)\tilde{t}+\tilde{r}]\psi_{1,-}\\
    (U\psi)_{0,-}&=\tilde{r}\psi_{-1,+}+d\tilde{t}\psi_{1,-}\\
    (U\psi)_{d,+}&=d\tilde{t}\psi_{d-1,+}\\
    (U\psi)_{d,-}&=[(d-1)\tilde{t}+\tilde{r}]\psi_{d-1,+}\\
    (U\psi)_{w,+}&=tw\psi_{w-1,+}+[r+(d-w-1)t]\psi_{w+1,-}\\
    (U\psi)_{w,-}&=t(d-w)\psi_{w+1,-}+[t(w-1)+r]\psi_{w-1,+}
  \end{split}
\end{equation}

We begin with the particle in the state $\ket{-1,0}$, \ie a particle
localized at the vertex just left of the hypercube on the tail, and
pointing to the right (see Fig.\ref{fig:potential}).

\addfigure[.55]{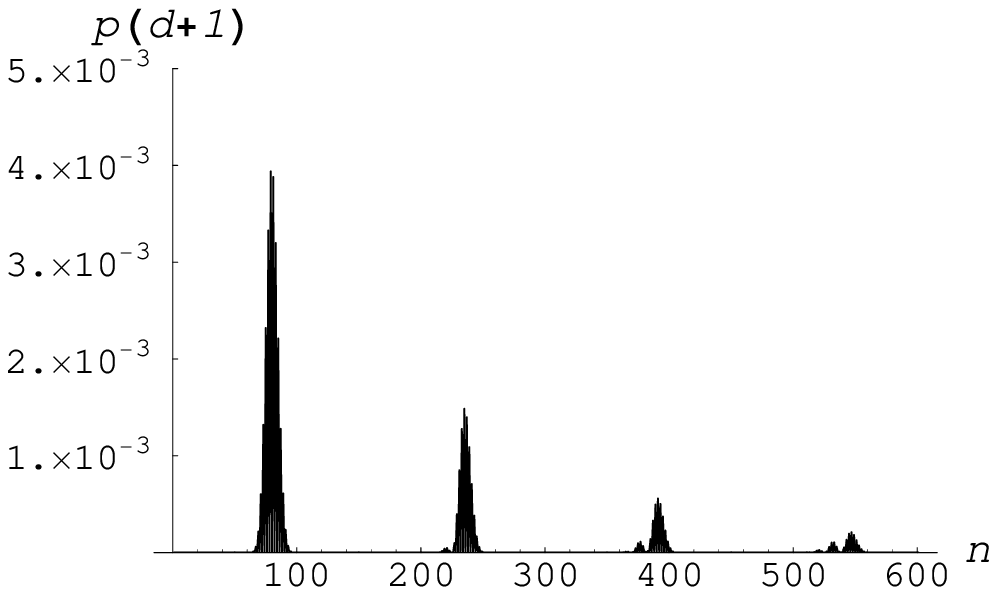}{The scattering probability of dim 10
  hypercube, for a photon incoming from the source (S) ($n$ is the number of
  steps). The multiports are symmetric.}

We have simulated the probability that a particle incoming from the left will
be absorbed by the detector after $n$ steps (see Fig. \ref{fig:probdist_g}). This
means that the system evolves for $n$ steps, and then a projection on
the basis $\{\ket{d,+}\}$ is made.
The result shows periodic beats of the probability of absorption of the photon by
the detector.

\section{Non-symmetric initial state}
\label{sec:nonsymmetric}
When we impose a symmetry condition on initial states the whole
problem becomes
linear in the dimension of the hypercube (normally its
complexity is exponential in $d$). But we also may be interested in
the behavior appears as a result of phase differences in the
components of the initial state. Given the $d$-hypercube with
semi-infinite tails attached to the vertices $0\dots0$ and $1\dots1$,
and the initial state
\begin{equation}
  \ket{\psi_0}=\sum_{j=1}^d\gamma_j\ket{0\dots0;j}
\end{equation}
Now the amplitude  of $U^d\ket{\psi_0}$ to project on
$\ket{1\dots1;+}$ is given by the sum of amplitudes to traverse from
$0\dots0$ to $1\dots1$ in $d$ steps, which is $t^{(d-1)}\tilde{t}$. For
initial state to be a concrete $\ket{0\dots0;j}$ we have $(d-1)!$
such paths. The situation is analogous for all $j$, with each initial
direction $j$ contributing the factor $\gamma_j$ to the amplitude. The
overall amplitude is
\begin{equation}
  \bra{1\dots1,+}U^d\ket{\psi_0}=\sum_{j=1}^d\gamma_j(d-1)!t^{(d-1)}\tilde{t}
\end{equation}
The probability of detecting the particle at the state
$\ket{1\dots1;+}$ depends only on $\sum_{j=1}^d\gamma_j$. In this
sense the hypercube with tails attached behaves like Mach-Zehnder
interferometer.




\section{Implementing the SQRW}
\label{sec:implementation}
Until now we haven't discussed the question whether it is feasible to
implement the SQRW. To build a whole network of multiports we need
exponentially growing resources (the number of vertices grows
exponentially). However, to encode the states we need only
$d\lceil\log{d}\rceil$ qubits. So we can ask a question: Is it possible to build a network of
quantum gates operating on the qubit register of this size? This is
most easily done only on the hypercube without the semi-infinite tails
attached; however, it is also possible to implement this scheme, by
adding some overhead of gates to the network.
We need $d$ qubits for the position register $\ket{x}$ and at least
$\lceil\log{d}\rceil$ qubits for the direction register $\ket{\phi}$.
The first part of one application of the unitary
operator $U$ is controlled negation of each bit of $x$ depending on
$\ket{\varphi}$. The second part is the transformation of the state
$\ket{\varphi}$, so that the action  of the multiports is accounted. More
precisely, the first part is
\begin{equation}
  \label{eq:gate1}
  \ket{x}\ket\varphi\rightarrow\sum_a\ket{x+a}\braket{a}{\varphi}\ket{a}\, ,
\end{equation}
and the second part reads
\begin{equation}
  \label{eq:gate2}
  \ket{x}\ket{a}\rightarrow\ket{x}\big[r\ket{a}+\sum_{b\neq a}t\ket{b}\big]\, ,
\end{equation}
for each $a=1,\dots,d$. The first part described by Eq.~(\ref{eq:gate1}) can be
implemented using variant of CNOT gate. The CNOT gate operates on two
qubits such that it negates the first (target) qubit, iff the second
(control) qubit is nonzero. In the symbolic form, we have:
$\text{CNOT}=\sigma_x\otimes\ket1\bra1+1\otimes\ket0\bra0$. We employ
the \pcnot gate, which differs from the CNOT gate in that it has a
$d$-dimensional control state, unlike a single qubit. If the control
state is $\ket\phi$ (the accepting control state),
then the target qubit is negated, otherwise,
it is kept in the original state. The operational form of the
\pcnot gate is
\begin{equation}
  \label{eq:phiCNOT}
  \pcnot=\sigma_x\otimes\ket\phi\bra\phi+1\otimes(1-\ket\phi\bra\phi)\, .
\end{equation}
Obviously, the \pcnot is unitary.

The operation (\ref{eq:gate1}) can be implemented using $d$
\pcnot gates (see Fig.\ref{fig:gate}). Each gate operates on a
different qubit from the position register. If the gate operates on
the $a$-th qubit, the accepting control state is chosen to be $\ket{a}$ (see
Fig.~\ref{fig:gate}).
\addfigure[.5]{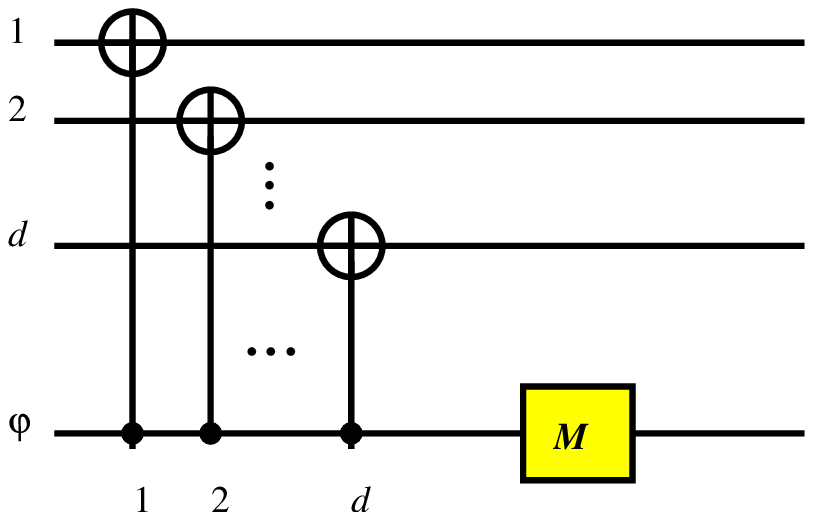}{The gate which implements the SQRW on the
  $d$-hypercube. The input state is the position register ($d$ qubits
  labeled $1,\dots, d$) and the direction register $\ket\varphi$.
  There are $d$ \pcnot gates stacked together, with accepting states
  $\ket{1},\dots,\ket{d}$, which change the position register, and the
  gate   $M$ which changes  the direction register.}

The operation (\ref{eq:gate2}) is implemented using a single unitary
operation $M$ operating on the direction  register. It corresponds to
the transformation of the state due to the multiports.
Its reads
\begin{equation}
  \label{eq:M}
  M=\sum_a\big[r\ket{a}\bra{a}+t\sum_{b\neq a}\ket{b}\bra{b}\big]\, .
\end{equation}
The $d\times d$ matrix form of $M$ is
\begin{equation}
  \label{eq:Mmatrix}
  M=
  \begin{pmatrix}
    r&t&\dots&t\\
    t&r&\dots&\\
    \vdots&&\ddots&\\
    t&\dots&t&r
  \end{pmatrix}
\end{equation}
Consequently, the unitary evolution operator $U$ of the SQRW on the
hypercube may be decomposed as $U=G_2G_1=(1\otimes M)C_1\dots C_d$,
where $C_a=\sigma_x\otimes\ket{a}\bra{a}+1\otimes(1-\ket{a}\bra{a})$
is the \pcnot operator with the target qubit being the $a$-th
qubit from the position register, and accepting the state  $\ket\phi=\ket{a}$, with
$a=1,\dots,d$. The operators $C_a$ are mutually commuting, and have
the common eigenvectors. To find them, we decompose the eigenvectors
into the product of $2\times d$ dimensional vectors
$\ket\psi\otimes\ket\chi$. Applying $C_a$ on $\ket\psi\ket\chi$  we obtain (providing
that $\sigma_x\ket\psi=\lambda\ket\psi$)
\begin{equation}
  \begin{split}
    C_a\ket\psi\ket\chi&=\sigma_x\ket\psi\ket{a}\braket{a}{\chi}+\ket\psi\otimes(\ket\chi-\ket{a}\braket{a}{\chi})\\
    &=\ket\psi\otimes[(\lambda-1)\braket{a}{\chi}\ket{a}+\ket\chi]\\
    &=
    \begin{cases}
      \ket\psi\ket\chi&:\lambda=1\, ;\\
      \ket\psi\otimes[\ket\chi-2\braket{a}{\chi}\ket{a}]&:\lambda=-1\, .
    \end{cases}
  \end{split}
\end{equation}
The case with $\lambda=-1$ has to be dealed separately. If
$\ket\chi=\ket{a}$, then we get $C_a\ket\psi\ket{a}=-\ket\psi\ket{a}$,
and if $\ket\chi=\ket{b},\;b\neq a$ then
$C_a\ket\psi\ket{b}=\ket\psi\ket{b}$. The conclusion: the eigensystem
of $C_a$ is the set of vectors $\ket\psi\ket\chi$, where $\ket\psi$ is
the eigenvector of $\sigma_x$ and $\chi=\ket{a},\;a=1,\dots,d$.
What about the eigensystem of $G_2$?
The matrix $M$ can be diagonalized, since it is translation
invariant. We search for the eigenvectors in the form
$\psi_k=\sum_{a=1,d}e^{2\pi ika/d}e_a$, which yields
\begin{equation}
  \begin{split}
    M\psi_k&=\sum_{a=1,d}\big(re^{2\pi ika/d}+\sum_{b=1,d-1}te^{2\pi
      ik(a\ominus b)/d}\big)e_a\\
    &=\underbrace{\big(r+t\sum_{b=1,d-1}e^{-2\pi ikb/d}\big)}_{\lambda_k}\sum_{a=1,d}e^{2\pi ika/d}e_a\, .
  \end{split}
\end{equation}
where $\lambda_k=r-t$, if $k\neq0$, and $\lambda_k=r-t+td$.

\section{SQRW is  superset of  coined quantum random walk}
\label{sec:superset}

In this section we will discuss the connection between the scattering and the coined quantum
random walks.
The SQRW reduces to coined quantum random walk
on a regular graph (having all vertices with the same degree), and
conversely, the SQRW is the generalization of the coined quantum
random walk on general graphs.

There is an isomorphism between coined quantum random walk (CQRW) and
SQRW on the same Cayley graph over Abelian group, $\mathcal{G}$. We
recall that CQRW is defined 
by an unitary operator $E$ on  Hilbert space
$\hilb_E=\hilb_X\otimes\hilb_A$ where $\hilb_X$ is spanned by vectors
$\ket{x},x\in\mathcal{G_V}$ and  $\hilb_A$ is spanned by the generators
of $\mathcal{G}$, the basis vectors $\ket{a}$. One step of CQRW is
given by $E=SC$ where $S=\sum_a T_a\otimes\pi_a$ and $C=1\otimes
M$. Here $T_a\ket{x}=\ket{x+a}$
is the translation, $\pi_a$ is the 
projection to $\ket{a}$ and $M$ is any unitary operator. The isometry
is given by 1-1 mapping of basis vectors of both $\hilb$ and $\hilb_E$
like $\ket{x(x+a)}_{\hilb}\equiv\ket{x}\ket{a}_{\hilb_E}$. The
correspondence between operators $U$ and $E$ is:  $U_x\ket{yx}$, where
$y+a=x$, is the same like applying the translation $S$ on
$\ket{x-a}\ket{a}$ and then applying the coin $M$ on $\ket{a}$
s.t. matrix representations of $U_x$ in natural basis of
$\hilb_x,\widetilde\hilb_x$ and $M$ in natural basis of $\hilb_A$ are
the same.

For regular graphs, we can decompose $\hilb_E$ into direct product of
$\hilb_{Ex}$ s.t. $\hilb_{Ex}=\text{span}\{\ket{x-a}\ket{a}:a \text{ is
  gen. }\mathcal{G}\}$.

The scheme for generalizing coined quantum random walk on general
graphs was proposed in \cite{kendon2003}, but this required an oracle
which operate on the set of all edges of the graph. Our scheme is
based on local operations done by multiports, so it is more reasonable
and easier to implement physically. This was actually proposed in
\cite{ambainis2004}. 
Algorithms based on coined
quantum random walks were proposed e.g. in
\cite{eisenberg2003}

\section{Searching with SQRW}
\label{sec:search}

In this section we will address a question whether
it is possible to use the SQRW for a database search, or a similar
task? To answer this question, we need to formulate what a quantum
database is and how we can move around its entries using the SQRW.

The database we are  searching in is the so called quantum dictionary. The
classical dictionary is a set of pairs (key, value). The set of all
keys is given by the  topology of a graph, yielding the adjacency
relations among all the keys. Random walk (classical) in the
dictionary is bound to the edges of this graph.

The searching problem in the dictionary is given as follows: given a
value, find a key, such that (key, value) is in the dictionary.
For $N$ keys, this is an $O(N)$ problem.
To obtain a quantum version of this scheme, we have to ``quantize'' (non-canonically)
the problem.
Due to the fact that the graph is not regular, we cannot
factorize the complete Hilbert space, but we need to label the states
in the most general fashion: $\ket{x,y}$, where $(x,y)$ is an edge.
The searching procedure consists of applying one step of the SQRW, and
then by querying the database. The query corresponds to an application  of a
unitary operator (the oracle) \cite{nielsen2000}, which flips some auxiliary qubit,
depending on whether the value assigned to the key is the one we are
searching for. That is, the oracle is the transformation
$\mathcal{O}\ket{x,y}\ket{q}\mapsto\ket{x,y}\ket{q\oplus f(x,y)}$, where $f(x,y)$
gives the value 1 if any of the vertices $x,y$ satisfy the query, and 0
otherwise. It is clear that the oracle $\mathcal{O}$ is unitary. Now
the searching algorithm is based on the sequence of operations
$(\mathcal{O}U)^n$, where $U$ makes one step of the SQRW and
$\mathcal{O}$ is the oracle query (equivalent to the action of the
multiports).
One such algorithm has been presented in Ref.~\cite{shenvi2003}: In our terms
it is the SQRW on the
hypercube, where the multiport assigned to one marked key has trivial
coefficients $r,t$ (they only change the phase), while the other
multiports have coefficients corresponding to the action of the Grover
operator to the direction states. In Ref.~\cite{shenvi2003} it has been shown that
the marked key can be found in $O(\sqrt{N})$ steps with probability $O(1)$,
where $N$ is the number of the vertices of the hypercube.

\section{Symmetries of the evolution operator $U$}

\label{sec:symmetries}

The basic relation $U\psi=\lambda\psi$, where
$\psi=\sum_{xa}\gamma_{xa}\ket{xa}$ yields the following
recurrence relation:
\begin{equation}
  \label{eq:rrel}
  r\gamma_{x,-a}+t\sum_{b\neq a}\gamma_{xb}=\lambda\gamma_{x+a,a}
\end{equation}
Finding the symmetries of this operator helps us to find its eigensystem.
We can Fourier transform the states of $\hilb$ to another basis, in which the
solutions can be  easier found. The operator $U$ has many symmetries,
one of them is the translation $T_b:x\mapsto x+b$.
The eigenvectors of $T_b$ are
(for details see Ref.~\cite{moore2001})
\begin{equation}
  \label{eq:fourier}
  \tilde{\ket{ka}}=\sum_x(-1)^{kx}\ket{xa}\, ,
\end{equation}
with eigenvalues $(-1)^{k_b},\;k\in\bbz_2^d$. The action of $U$ on
$\tilde{\ket{ka}}$ is
\begin{equation}
  \label{eq:u_on_ka}
  U\tilde{\ket{ka}}=(-1)^{k_a}\Big(r\tilde{\ket{ka}}+\sum_{b\neq
    a}\tilde{\ket{kb}}\Big)\, ,
\end{equation}
and in the basis $\tilde{\ket{ka}}$ $U$ has the form
$\tilde{U}=\text{diag}\,(\{\tilde{V}_k\})$ where
\begin{equation}
  \label{eq:tilde_v_k}
  \tilde{V}_k=
  \begin{pmatrix}
    r(-1)^{k_1}&t(-1)^{k_2}&\dots&t(-1)^{k_d}\\
    t(-1)^{k_1}&r(-1)^{k_2}&t(-1)^{k_3}&\dots\\
    &\vdots&\ddots&\dots\\
    t(-1)^{k_1}&\dots&&r(-1)^{k_d}
  \end{pmatrix}\, .
\end{equation}
Now we only  need to find the eigensystem of this comparatively small
matrix. It is obvious, that $\tilde{V}_k$ is translationally symmetric.
The eigensystem of (\ref{eq:tilde_v_k}) can be found in Ref.~\cite{moore2001}.

In what follows we will find
another symmetry. Unlike the previous case, now we will be
changing both the elements of the position and the direction Hilbert
spaces. This transformation $R$ will change the vector $\ket{x,a}$
s.t. the binary string $x$ is cyclically shifted right by one place
and $a$ is set to $a\oplus1$ modulo $d$. Since $a$ is unambiguously
defined by the position in the binary string at which $x$ differs from
$x+a$, this transformation is a symmetry. This can be viewed of as a
rotation about the line segment connecting two opposite vertices
$0\dots0$ and $1\dots1$. We can choose any other two vertices $x,y$
s.t. $|x-y|=d$, and get a symmetry operator $R_{xy}=B_x^\dagger RB_x$,
where $B_x$ changes the role of $0$ to $x$ and $1$ to $1+x$. More
precisely, $B_x\ket{z,a}=\ket{z+x,a}$ (hence $B^\dagger=B$).
Two transformations $R_x,R_y$ generally do not commute, but they both
commute with $U$.

\section{Conclusion}
We have proved that
the SQRW is in fact a version of the coined
quantum random walk. We can use this observation to extend the coined quantum random
walk to the cases of non-regular graphs. While it is in principle easy
to construct the SQRW on any graph, it is still a question whether we also
can simulate it efficiently (e.g. like in
Sec.\ref{sec:implementation}). This point is crucial for  further
development of  quantum algorithms based on the SQRW in higher
dimensions (where the speedup may become noticeable). The class of
algorithms based on SQRW is the database searching, using the oracle
queries along with the ``random'' steps. We already know at least one such
algorithm (see Ref.~\cite{shenvi2003}) and we know that it is optimal. We cannot
expect the complexity drop below $O(\sqrt{N})$ for $N$ database keys,
but the new algorithms may be more general in their inputs, and maybe
more easy to implement.

We have found the connection between mixing properties of the
multiport (or the coin), and the distance of the respective operator
from the unity. It might be interesting to find an exact function of this
distance, which yields the measure of mixing for the SQRW.

\begin{acknowledgments}
  This work was supported by the European Union  projects
  QUPRODIS, QGATES and CONQUEST.
\end{acknowledgments}



\begin{thebibliography}{99}
\bibitem{footnote1}
  We should note that there is nothing
  really {\em random} (in a classical sense) about quantum random walk:
  the quantum evolution is always
  unitary (deterministic), and the only randomness involved in the scheme are the random
  outcomes of the measurements.
\end{thebibliography}
\end{document}